\documentclass{elsart}
\usepackage{epsfig,graphicx,amsfonts,amssymb,color}
\begin{document}
\begin{frontmatter}
\title{Nonextensive statistical features of the Polish stock 
market fluctuations}

\author{R. Rak$^{1}$, S.~Dro\.zd\.z$^{1,2}$, J.~Kwapie\'n$^2$}

\address{$^1$ Institute of Physics, University
of Rzesz\'ow, PL--35-959 Rzesz\'ow, Poland}
\address{$^2$Institute of Nuclear Physics, Polish Academy of Sciences,
PL--31-342 Krak\'ow, Poland}

\begin{abstract}
The statistics of return distributions on various time scales
constitutes one of the most informative characteristics of the
financial dynamics. Here we present a systematic study of such
characteristics for the Polish stock market index WIG20 over the
period 04.01.1999--31.10.2005 for the time lags ranging from one
minute up to one hour. This market is commonly classified as
emerging. Still on the shortest time scales studied we find that
the tails of the return distributions are consistent with the
inverse cubic power-law, as identified previously for majority of
the mature markets. Within the time scales studied a quick and
considerable departure from this law towards a Gaussian can
however be traced. Interestingly, all the forms of the
distributions observed can be comprised by the single
$q$-Gaussians which provide a satisfactory and at the same time
compact representation of the distribution of return fluctuations
over all magnitudes of their variation. The corresponding
nonextensivity parameter $q$ is found to systematically decrease
when increasing the time scales. The temporal correlations quantified 
here in terms of multifractality provide further arguments in favour 
of nonextensivity.
\end{abstract}

\end{frontmatter}

\section{Introduction}

Making the quantification of financial fluctuations is a real
interdisciplinary challenge. The related well identified stylised
fact is the so-called inverse cubic power-law~\cite{Gabaix} which
applies to developed stock markets~\cite{Lux,Gopi,Plerou,Drozdz1},
to the commodity market~\cite{Matia6}, as well as to the most
traded currency exchange rates~\cite{Muller}. The emerging stock
markets are commonly considered to be governed by a somewhat
different dynamics which often~\cite{Matia,Miranda} results in
exponential tails of the return distributions. Of course, both the
above types of distributions are L\'evy unstable and thus for the
sufficiently long time lags they may converge towards a Gaussian.
The distribution with an exponential tail might correspond to an
intermediate stage between a distribution with the power-law
asymptotics and a very large time lag limit - a
Gaussian~\cite{Silva}. Such a scenario corresponds for instance to
the Heston model~\cite{Heston}.

In order to elaborate more on this sort of issues we
systematically study the character of fluctuations of the Polish
stock market as represented by the WIG20 index. This equity market
started trading on April 16, 1991 and the presently most often
quoted corresponding index is WIG20 (Warszawski Index Gie{\l}dowy
- Warsaw Stock Market Index), introduced in 1994, comprising
capitalization weighted prices of the 20 largest companies. The
high quality electronic processing and recording of all the
transactions started in the beginning of 1999. The analysis
presented here thus covers the time period since January 4, 1999
until October 31, 2005.

\begin{figure}
\hspace{-0.2cm}
\epsfxsize 14.0cm
\epsffile{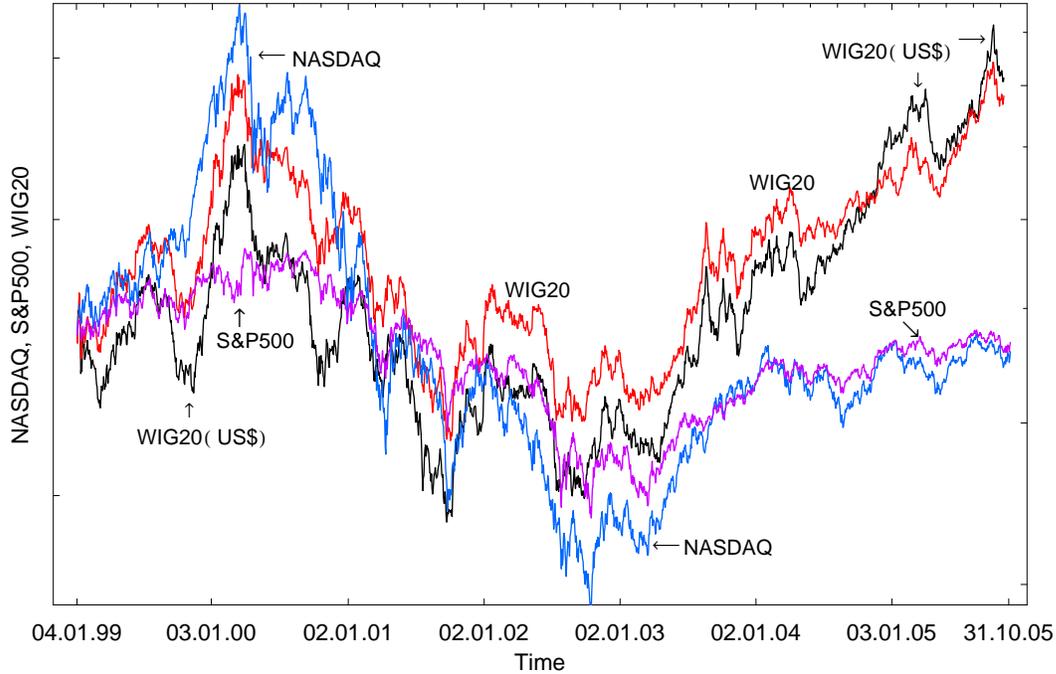}
\caption{NASDAQ, S\&P500 and WIG20(Warsaw Stock Exchange) index
from 1999.01.04 until 2005.10.31.}
\end{figure}

The daily trading closing hour during this period was 4:00pm and since
17.11.2000 4:10pm. The opening hours have been changed two times. On
4.1.1999 until 30.7.1999 (Period1) it was 1pm, then until 16.11.2000
(Period2) it was 12am, and then 10am (Period3) with the closing at
4:10pm.\\
As it can be seen from Fig.~1 in the whole time period inspected
here, even though representing an emerging market, the WIG20 has
been following the overall world trend - more in terms of the
phase than in the amplitude however. During the first two years of
the period considered its behaviour closely resembles the Nasdaq.
Starting in 2003 it however by far overperforms the two world
major indices: the Nasdaq and the S$\&$P500. As natural, the
original WIG20 is quoted in the Polish Zloty (PLN). Converting
systematically the PLN into the US$\$$ - to make this comparison
even more informative - results in an even larger gain as can be
easily seen from Fig.~1. This is due to a parallel sizable PLN
appreciation in the period considered.

\section{Conventional log-log analysis}

For the time series $W(t)$ representing the index value at time
$t$ we use the commonly accepted definition of returns as
\begin{equation}
R \equiv R(t,\Delta t) = \ln W(t + \Delta t) - \ln W(t). \label{G}
\end{equation}
As another standard procedure, we calculate the normalized returns
$r \equiv r(t,\Delta t)$ defined as
\begin{equation}
r = {R - \langle R \rangle_T \over v}, \label{g}
\end{equation}
where $v \equiv v(\Delta t)$ is the standard deviation of returns
over the period $T$
\begin{equation}
v^2 = \langle R^2 \rangle_T - \langle R \rangle_T^2 \label{v}
\end{equation}
and $\langle \dots \rangle_T$ denotes a time average.

The cumulative distribution functional (cdf) of $\Delta t=1$min
moduli of WIG20 returns collected from the whole period 1999 -
2005 specified above is shown in panel (a) of Fig.~2.
Interestingly, this distribution displays very similar behaviour as
for many mature markets analysed before. Even the tails of this
distribution reveal scaling
\begin{equation}
P(r > x) \sim x^{- \alpha}, \label{P}
\end{equation}
consistent in addition with the inverse cubic power law
($\alpha=3$). The remaining (b), (c) and (d) panels show cdf's
separately for the Period1, Period2 and Period3. As one can see,
even though these periods correspond to different phases - from
less to more advanced - of the Polish Stock Market development,
the return distribution characteristics remain essentially
invariant. This indicates that the fluctuation characteristics of
an emerging market do not have to differ from those of a mature
one.
\begin{figure}
\hspace{-0.4cm}
\epsfxsize 14.5cm
\epsffile{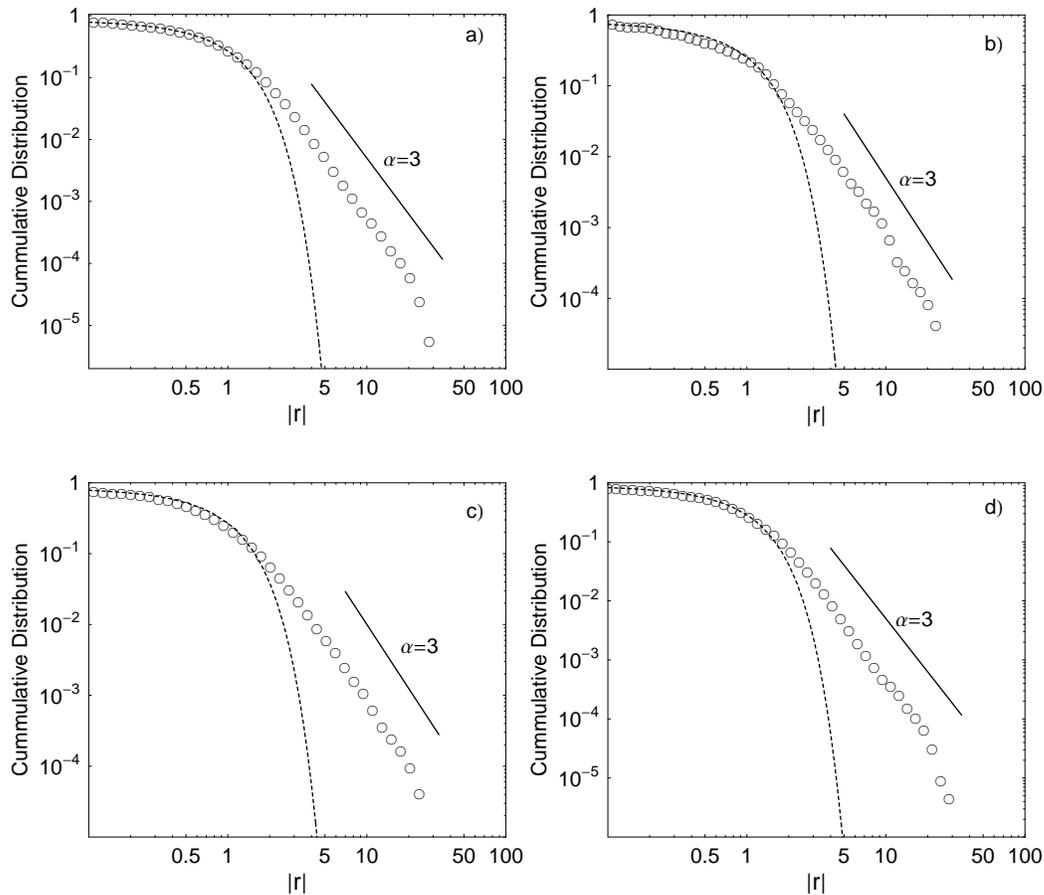} 
\caption{Cumulative distributions of moduli of the WIG20 normalised returns
( $\Delta t= 1$ min, ({\scriptsize $\bigcirc$})) for four periods:
a) the whole period 4 Jan 1999 $-$ 31 Oct 2005; b) Period1 from 4
Jan 1999 to 30 Jul 1999; c) Period2 from 2 Aug 1999 to 16 Nov
2000; d) Period3 from 17 Nov 2000 to 31 Oct 2005. Dashed line
corresponds to a Gaussian distribution.}
\end{figure}
A distribution whose tails follow the inverse cubic power law has
a finite second moment and is thus L\'evy unstable. In the present
context this means that the return distributions for the
sufficiently large time lags $\Delta t$ are expected to eventually
start converging towards a Gaussian due to effects in the spirit
of the Central Limit Theorem (CLT). Of course, fluctuations
typically carry some higher order time correlations - quantifiable
for instance in terms of multifractals~\cite{Oswiecimka,Kwapien} -
therefore this convergence may be much slower than for
uncorrelated stochastic processes to which the conventional CLT
refers. For the stock market fluctuations such effects are however
identified in the literature~\cite{Gopi,Plerou} for the time
scales of the order of a few days or even a few hours for the more
recent data~\cite{Drozdz1}. How the corresponding situation
develops for our WIG20 data when the time lag $\Delta t$ increases
is shown in Fig.~3.\\Departure from the inverse cubic scaling can
already be seen for time lags larger than 10 minutes and for
$\Delta t=60$ min the scaling regime is hardly visible with the
tail of the distribution being closer to a Gaussian.

\begin{figure}
\hspace{0.6cm}
\epsfxsize 12.5cm
\epsffile{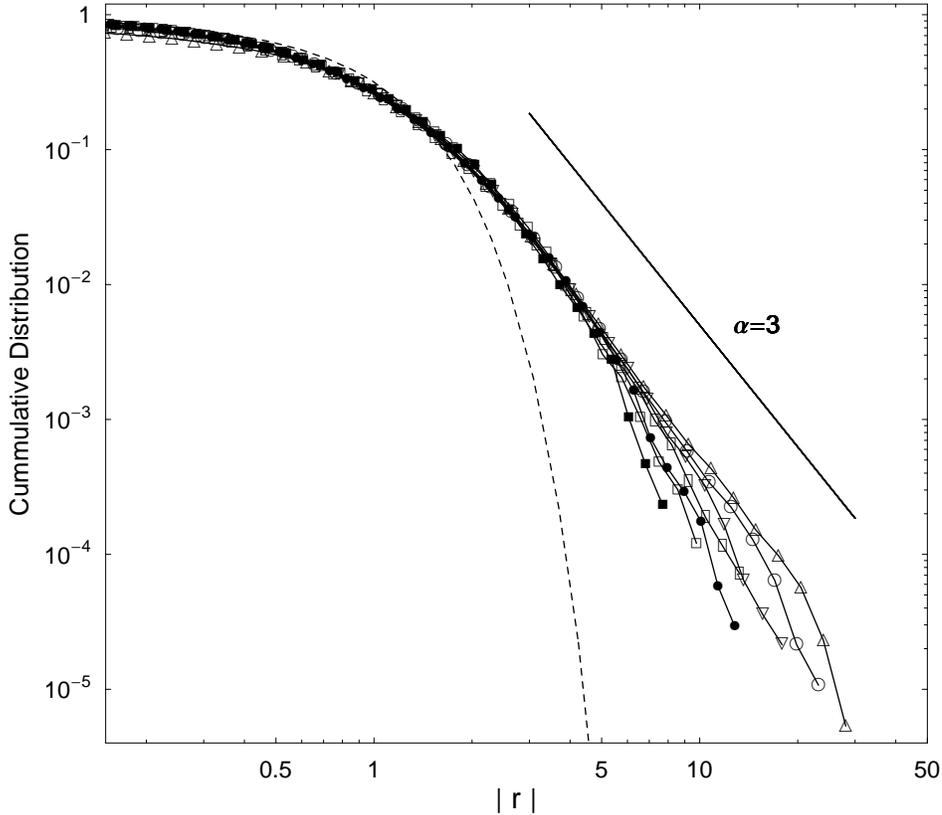}
\caption{Cumulative distributions of moduli of the WIG20 normalized returns
(symbols) from 4 Jan 1999 to 31 Oct 2005 for several time scales: 1 min ({\scriptsize 
$\bigtriangleup$}), 2 min ({\scriptsize $\bigcirc$}), 4 min ({\scriptsize 
$\bigtriangledown$}), 8 min ({\scriptsize $\square$}), 16 min ({\scriptsize $\bullet$}), 32 
min ({\scriptsize $\square$}), and 60 min ({\scriptsize $\blacksquare$}). Dashed line 
corresponds to a Gaussian distribution.}
\end{figure}

Whether this much faster convergence with $\Delta t$ towards a normal distribution is 
characteristic to WIG20 fluctuations or it just reflects the fact that one deals here with 
even more recent data than in ref.~\cite{Drozdz1} remains an open question. Based on 
empirical arguments collected form the world leading stock markets the hypothesis put forward 
in that reference says that when going from past to present the same $\Delta t$ measured by a 
conventional clock time effectively corresponds to increasing time lags of an internal market 
time and this originates form an increasing speed of the information processing.

\section{Nonextensive statistical approach}

The fat tails in the financial return distributions and the complex character of the 
underlying temporal correlations~\cite{Oswiecimka,Kwapien,Drozdz2,Bartolozzi} indicate that 
the conventional concept of ergodicity may break down in the financial dynamics. Under such 
circumstances the generalised formalism of nonextensive statistical mechanics may offer an 
appropriate framework to quantify the corresponding statistics. At present the most 
consistent seems the one based on the generalised entropy which for a set of $N$ events $\{ 
x_i\}$ characterised by the probabilities $\{p_i\}$ reads
\begin{equation}
S_q=-\sum\limits_{i=1}^{N} p_i^q \ln_q p_i, \label{Sq}
\end{equation}
as postulated by Tsallis~\cite{Tsallis}. Here $\ln_q$ denotes the $q$-logarithm function 
$$\ln_{q}x=(x^{1-q} - 1)/ (1-q).$$ The parameter $q$ in Eq.~\ref{Sq} is the so called 
nonextensivity parameter. For $q=1$ this equation expresses the standard Bolzmann-Gibbs 
entropy.

The optimization of this generalized entropic form under appropriate 
constraints~\cite{Tsallis,Tsallis1}, in the continuous form, yields the following 
$q$-Gaussian form for the distribution of probabilities
\begin{equation}
p\left( x\right) =\mathcal{N}_{q}\,e_{q}^{-\mathcal{B}_{q}\left( x-\bar{\mu}%
_{q}\right) ^{2}}\label{px}
\end{equation}
where
$$
\mathcal{N}_{q}=\left\{
\begin{array}{ccc}
\frac{\Gamma \left[ \frac{5-3q}{2-2q}\right] }{\Gamma \left[
\frac{2-q}{1-q}
\right] }\sqrt{\frac{1-q}{\pi }\mathcal{B}_{q}} & ~~for  & q<1 \\[3mm]
\frac{\Gamma \left(\frac{1}{q-1}\right)}{\Gamma \left(\frac{3-q}{2
(q-1)}\right) \sqrt{\frac{\pi}{(q-1) \mathcal{B}_q}}} & ~~for &
1<q<3
\end{array}
\right. ,$$
$$ \bar{\mu} _{q}= \,\int x\frac{\ \left[ p\left( x\right)
\right] ^{q}}{\int \left[ p\left( x\right) \right] ^{q}dx}\
dx\equiv \left\langle x\right\rangle _{q} ,
$$
$$
\mathcal{B}_{q}=\left[ \left( 3-1\right)
\,\bar{\sigma}_{q}^{2}\right] ^{-1}
$$
and $e_q^x$ denotes the $q$-exponential function defined as
\begin{equation}
e_{q}^{x}=\left[ 1+\left( 1-q\right) \,x\right] ^{ \frac{1}{1-q}}.
\label{eq}
\end{equation}

Another argument which makes this distribution attractive from the
present perspective is that for $q>1$ it asymptotically $(x>>1)$
develops a power law form $p(x)\sim x^{\frac{2}{1-q}}$. In
particular, for $q=3/2$, on the level of the cumulative
distribution, it recovers the inverse cubic power law. This is an
especially nice aspect of the functional form expressed by the
Eq.~\ref{px} because it at the same time provides a compact form
for the probability distribution for any value of $x$. Indeed, the
first attempts~\cite{Tsallis2} of the applicability of this form
to describe the probability distributions of the financial
fluctuations turn out quite promising. For all these reasons in
the following we explore a possibility to describe the empirical
WIG20 return distributions presented in Fig.~3 by a family of the
$q$-Gaussians. In order however to attain a better stability of
this analysis, instead of directly using the Eq.~\ref{px} we
convert it to the cumulative form by defining
\begin{equation}
P_{\pm}(x) =\mp\int_{\pm \infty}^x p(x')dx' \label{Pc}
\end{equation}
where the + and - signs correspond to the right and left wings of
the distribution correspondingly. By using here the Eq.~\ref{px}
one obtains
\begin{equation}
P_{\pm}(x) =\mathcal{N}_q\left(\frac{\sqrt{\pi } ~\Gamma
\left(\frac{1}{2} (3-q) ~\beta \right)}{2 ~\Gamma (\beta )~
\sqrt{\frac{\mathcal{B}_q}{\beta }}}\pm(x-\bar{\mu}_{q}) \,
   _2F_1(\alpha ,\beta ;\gamma ;\delta )\right)~,\label{Pcx}
\end{equation}
where,\\
$\alpha=\frac{1}{2}$, $\beta =\frac{1}{q-1}$, $\gamma
=\frac{3}{2}$, $\delta =-\mathcal{B}_q(q-1)(\bar{\mu}_{q}-x)^2$
and $_2F_1(\alpha ,\beta ;\gamma ;\delta)$ is the Gauss
hypergeometric function defined by the following power series
expansion:

$$
_2F_1(\alpha ,\beta ;\gamma ;\delta)=1+\frac{\alpha ~\beta}{1!
~\gamma } ~\delta+\frac{\alpha(\alpha +1) ~\beta(\beta +1)}{2!
~\gamma (\gamma+1)} ~\delta ^2+ ~\dots ~= \sum_{k=0}^\infty
{\frac{\delta ^k ~(\alpha)_k~(\beta)_k}{k!~(\gamma) _k}}
$$

\begin{figure}
\label{figx}
\hspace{1.2cm}
\epsfxsize 11.0cm
\epsffile{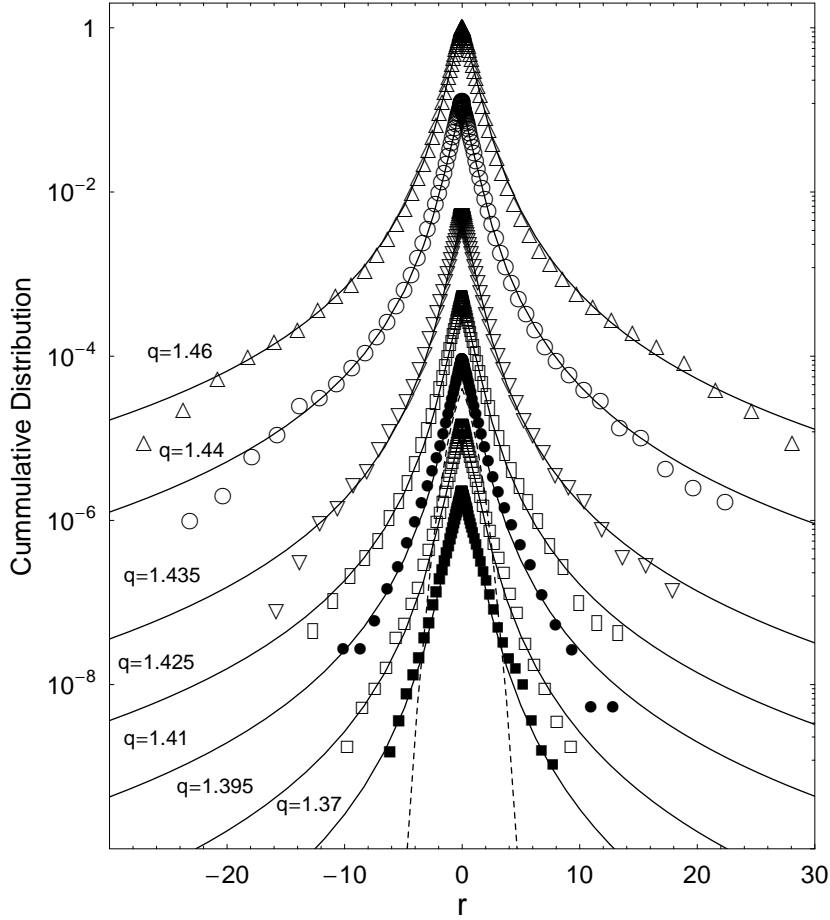}
\caption{Cumulative distributions of normalized returns (symbols)
for WIG20 index (Warsaw Stock Exchange) from 4 Jan 1999 to 31 Oct
2005 for several time scales. Solid line represents the best
theoretical fit corresponding to Eq.~\ref{Pcx}, from top to
bottom: $q=1.46$ ($\Delta t=1$ min, {\scriptsize
$\bigtriangleup$}), $q=1.44$ ($\Delta t=2$ min, {\scriptsize$
\bigcirc$}), $q=1.435$ ($\Delta t=4$ min,
{\scriptsize$\bigtriangledown$}), $q=1.425$ ($\Delta t=8$ min,
{\scriptsize$\square$}), $q=1.41$ ($\Delta t=16$ min, {\scriptsize
$\bullet$}), $q=1.395$ ($\Delta t=32$ min, {\scriptsize
$\square$}), and $q=1.37$ ($\Delta t=60$ min, {\scriptsize
$\blacksquare$}). Dashed line corresponds to a Gaussian
distribution. The curves have been rescaled vertically for better
display.}
\end{figure}

Fig.~4 shows the cumulative variant of the WIG20 data points for the same sequence of the 
time lags as in Fig.~3 and the corresponding best theoretical fits in terms of Eq.~\ref{Pcx}. 
The result appears very encouraging. For a given $\Delta t$ one obtains a good theoretical 
representation for the empirical probability distribution over the whole interval of changes 
of the returns. The only inaccuracy is at small $\Delta t$ for a few positive and even more 
negative extreme events whose probability is somewhat lower than what the overall global fit 
provides. Nevertheless, the obtained $q$-values for the smallest $\Delta t$ are close to 3/2, 
as consistent with the inverse cubic power law. With increasing $\Delta t$ the best fit 
$q$-values systematically decrease and the corresponding $q$-Gaussians provide amazingly 
reasonable representation for the empirical data on all the time scales considered.

\section{Temporal correlations}

Long-range, both space and time correlations, are the most characteristic features of 
the complex and nonextensive systems. Such interactions can manifest themselves as nonlinear 
correlations observed in empirical signals. In financial data these higher-order correlations 
can lead to some well-known stylized facts like, e.g., the multifractality, the persistent 
memory in volatility, the leverage effect and the fat tails of the distributions of returns. 
For the Polish stock market data, the latter of these stylized facts was already discussed in 
preceding sections and here we present results for the former ones.

\begin{figure}
\hspace{1.0cm}   
\epsfxsize 12.0cm
\epsffile{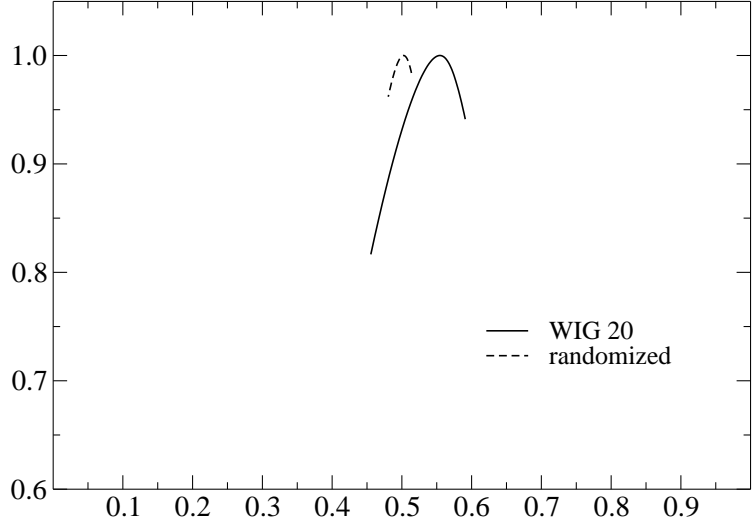}
\caption{Singularity spectra $f(\alpha)$ for original (solid line) and 
randomized (dashed line) time series of 1 minute WIG20 returns. The 
spectrum for randomized data was averaged over 100 independent 
generations.}
\end{figure}

We start with calculation of the singularity spectra $f(\alpha)$ that 
allows us to quantify the fractal properties of the data. We use the 
well-known method of multifractal detrended fluctuation analysis (MFDFA) 
which, according to our experience, gives the most reliable 
outcomes~\cite{oswiecimka06}. Technically, MF-DFA can be briefly sketched 
as follows~\cite{kantelhardt02}. For the time series $G$ of the returns 
$g(i), i=1,...,N$ one calculates the signal profile
\begin{equation}
Y(i) = \sum_{k=1}^i{(g(k)-<g>)}, \ i = 1,...,N
\end{equation}
where $<...>$ denotes the 
time-average of $G$. Now $Y(i)$ is divided into $M$ disjoint segments of 
length $n$ starting from the beginning of $G$ and $M$ equivalent segments 
starting from the end of $G$. For each segment $\nu, \nu=1,...,2M$, the 
local trend is to be calculated by least-squares fitting the polynomial 
$P_{\nu}^{(l)}$ of order $l$ to the data, and then the variance
\begin{equation}
F^2(\nu,n) = \frac{1}{n} \sum_{j=1}^n \{Y[(\nu-1) n+j] -
P_{\nu}^{(l)}(j)\}^2.
\label{std.dev}
\end{equation}
For the financial data the polynomial order as low as $l=2$ can be used. 
The variances (\ref{std.dev}) have to be averaged over all the segments 
$\nu$ and finally one gets the $q$th order fluctuation function
\begin{equation}
F_q(n) = \bigg\{ \frac{1}{2 M_s} \sum_{\nu=1}^{2 M_s} [F^2(\nu,n)]^{q/2}
\bigg\}^{1/q}, \ \ q \in \mathbf{R}.
\end{equation} 
The function $F_q(n)$ must be calculated for many different segments of 
lengths $n$. If the signal is fractal, the fluctuation function reveals 
power-law scaling
\begin{equation}
F_q(n) \sim n^{h(q)}
\label{scaling}
\end{equation}
for large $n$. The family of the scaling exponents $h(q)$ (the generalized 
Hurst exponents) can be then obtained by observing the slope of log-log 
plots of $F_q$ vs.~$n$. If $h(q)={\rm const}$ then the signal under study 
is monofractal; it is multifractal otherwise.  From the spectrum of 
the generalized Hurst exponents, one can calculate the singularity 
strength $\alpha$ and the singularity spectrum $f(\alpha)$ using the 
following relations (e.g.~\cite{kantelhardt02}):
\begin{equation}
\alpha=h(q)+q h'(q) \hspace{1.0cm} f(\alpha)=q [\alpha-h(q)] + 1,
\label{singularity}
\end{equation}
where $h'(q)$ denotes the derivative of $h(q)$ with respect to $q$.

\begin{figure}
\hspace{2.0cm}
\epsfxsize 10.0cm
\epsffile{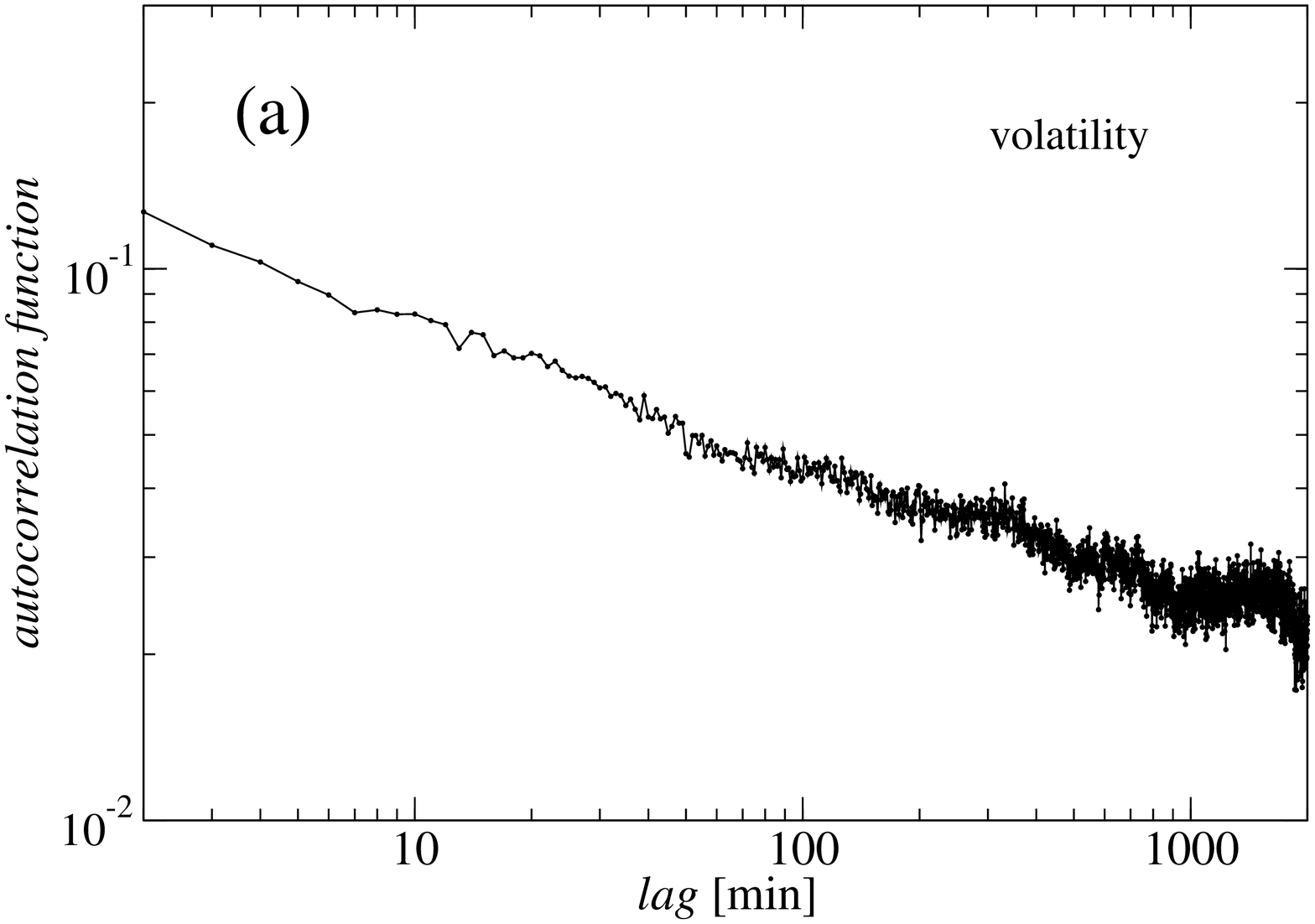}

\vspace{-0.8cm}
\hspace{2.0cm}
\epsfxsize 10.0cm
\epsffile{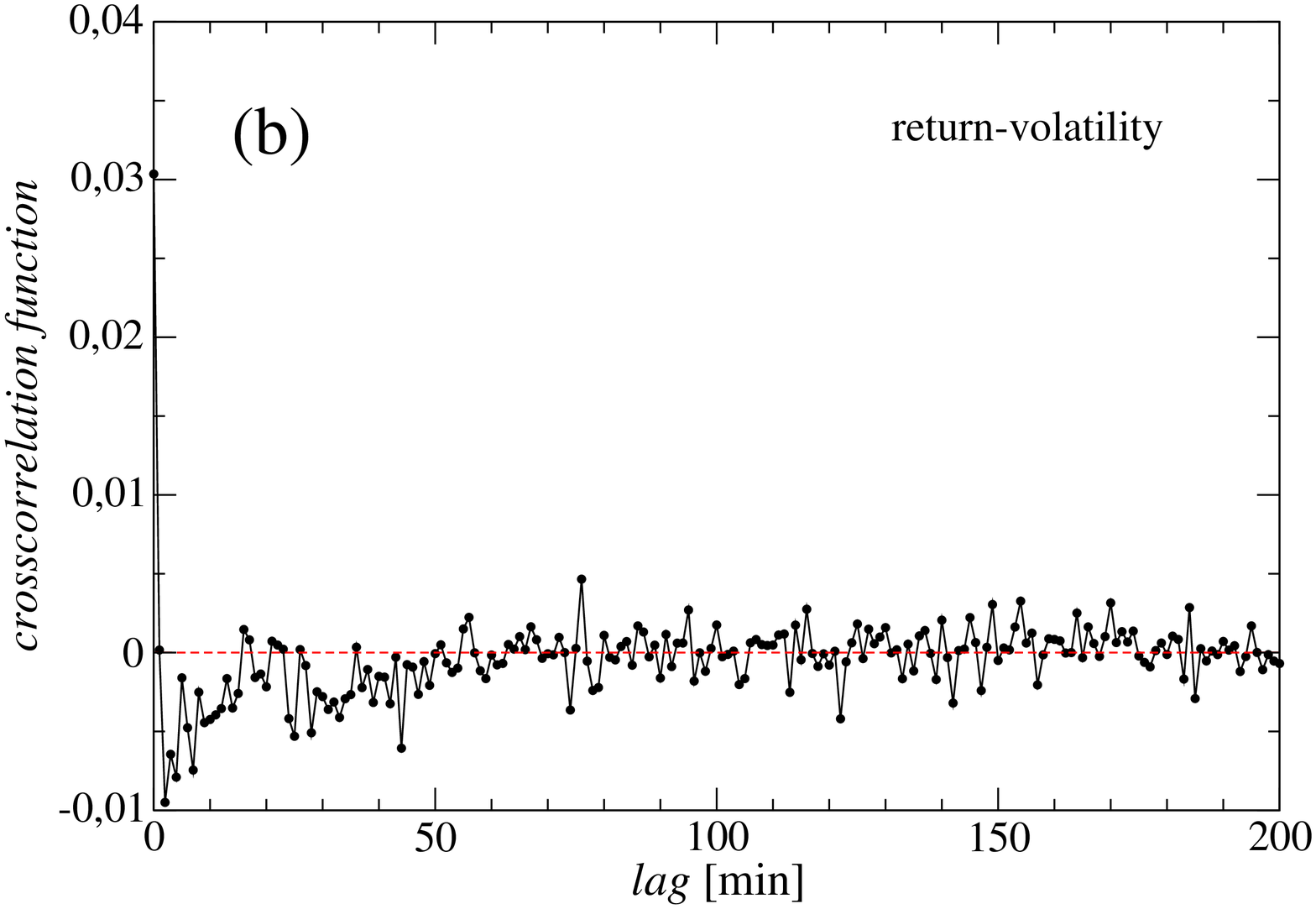}
\caption{(a) Volatility autocorrelation calculated for 1 min WIG20 returns
after removing the daily pattern and (b) crosscorrelation between 1  
minute returns and volatility (b). Zero level in (b) is denoted by dashed 
line.}
\end{figure}

Fig.~5 shows the singularity spectrum for the high-frequency WIG20 returns with $\Delta t=1$ 
min (solid line). On this time scale the number of data points is sufficiently large to allow 
a reliable study of this kind~\cite{oswiecimka06}. From the shape of $f(\alpha)$ curve it is 
evident that the signal under study has multifractal properties. This statement can receive 
an additional support from the shape of the spectrum for randomized data. The average 
spectrum for 100 independent realizations of the randomized data is also displayed in Fig.~5 
(dashed line). This spectrum differs substantially from the original one and is much closer 
to monofractal than the multifractal spectrum of the original series. As expected, the 
randomized data reveal no linear correlations and the spectrum is located at 0.5 which is in 
contrast with the spectrum for the original data showing a visible trace of such 
correlations. This example shows again~\cite{Kwapien} that these are the time correlations 
that constitute the main source of multifractality in the stock market dynamics.

The $f(\alpha)$ spectrum for our data indicates the existence of the correlations of 
different types; two basic ones are shown in Fig.~6. The volatility autocorrelation 
(Fig.6(a)) looks typically with its long temporal decay resembling its counterparts for the 
developed markets like, e.g., the American and the German ones~\cite{liu99,Drozdz1}. Also the 
negative crosscorrelation between the returns and the volatility (the leverage effect, 
Fig.~6(b)) looks similar to other data~\cite{perello,kwapien06}.

On the other hand, standard deviation of the returns (i.e., the time-averaged volatility) 
presented in Fig.~7 as a function of $\Delta t$ shows a somehow distinct behaviour than that 
observed for other markets~\cite{Gopi,Drozdz1}. Actually, for the first half of the 
considered time interval we observe three different scaling regions (see Fig.~7(a)): 
subdiffusive (scaling exponent $\delta = 0.45$) for the smallest time scales ($\Delta t \le 
3$ min), superdiffusive ($\delta = 0.60$) for medium time scales ($4 \le \Delta t \le 56$ 
min), and the region of approximately normal diffusion ($\delta = 0.51$) for the longest time 
scales ($\Delta t \ge 64$ min). For the second half of the considered time interval, 
corresponding to newer data, this scheme noticeably changes (Fig.~7(b)), mainly due to the 
broadening of the normal diffusion region which now covers $\Delta t > 10$ min, and the 
shortening of the now hardly identifiable weak superdiffusive region ($ 3 < \Delta t < 10$ 
min) with $\delta = 0.52$. This effect resembles the one observed for the American 
market~\cite{Gopi,Drozdz1} where the crossover point between the superdiffusion regime and 
the normal diffusion regime is shifted towards shorter time scales when going from past to 
present. It is noteworthy that this result goes in parallel with a faster convergence to the 
normal distribution of the return fluctuations observed in the previous section.

An interesting feature of the Polish stock market is the existence of the subdiffusive region 
for short time scales. This effect can be related to the antipersistence property of WIG20 
fluctuations on minute time scales. However, it remains unclear whether this antipersistence 
is unique to WIG20 or there is similar behaviour of other stock indices like, e.g., DJIA or 
DAX30, likely on time scales even shorter than 1 minute due to more sizeable volume traded in 
those latter cases. In parallel with the superdiffusion region, this subdiffusive regime in 
Fig.~7(b) weakens its character towards normal diffusion with $\delta = 0.48$. Only the 
crossover point between the subdiffusive and the superdiffusive region remains unchanged.

Since the scaling and multifractal properties of a system are associated with the long-range 
correlations, our results presented in this Section can be considered as yet another argument 
supporting the description of financial data in terms of the nonextensive statistical 
mechanics.

\begin{figure}
\hspace{2.0cm}
\epsfxsize 10.0cm
\epsffile{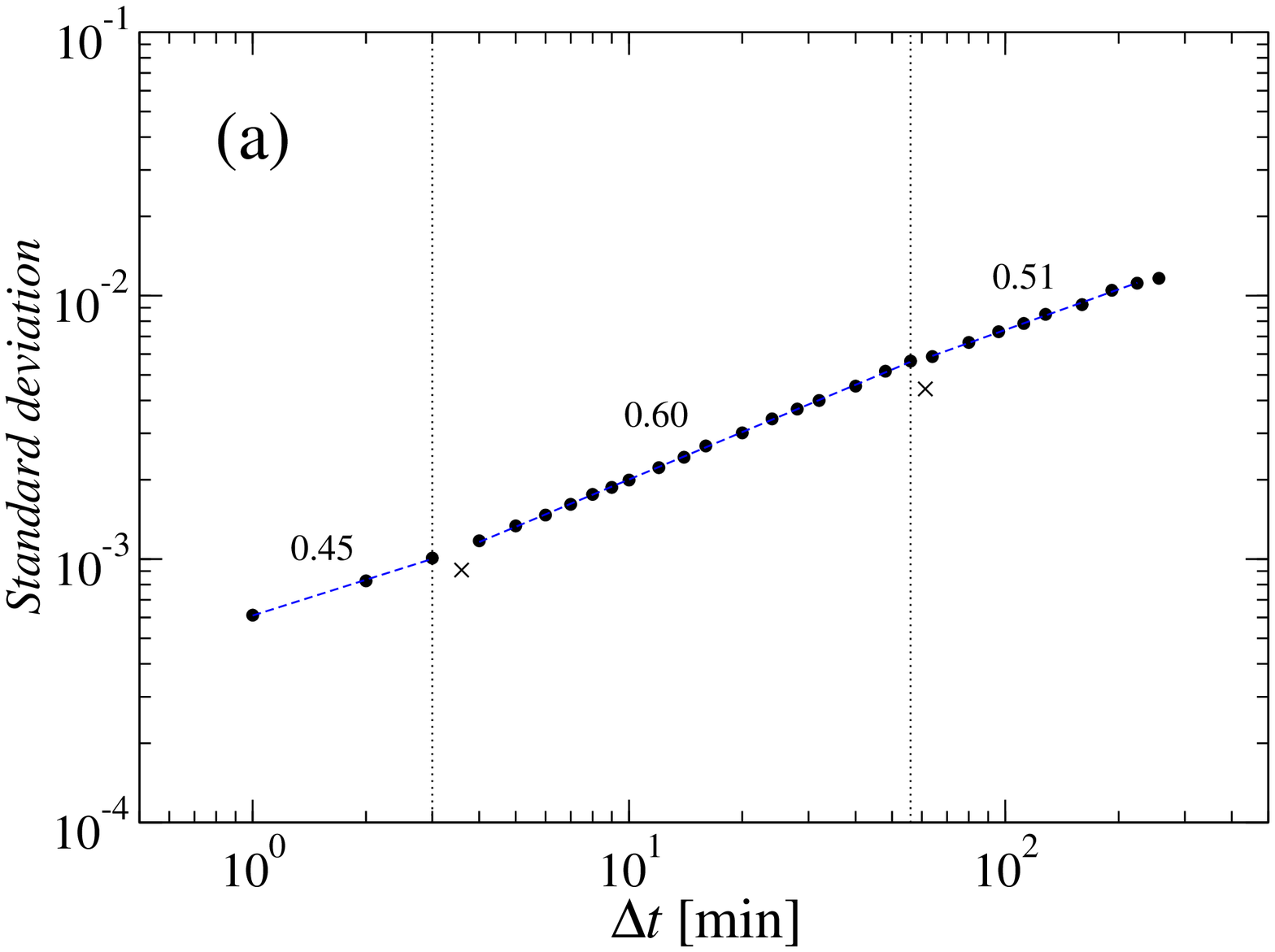}

\vspace{-0.8cm}
\hspace{2.0cm}
\epsfxsize 10.0cm
\epsffile{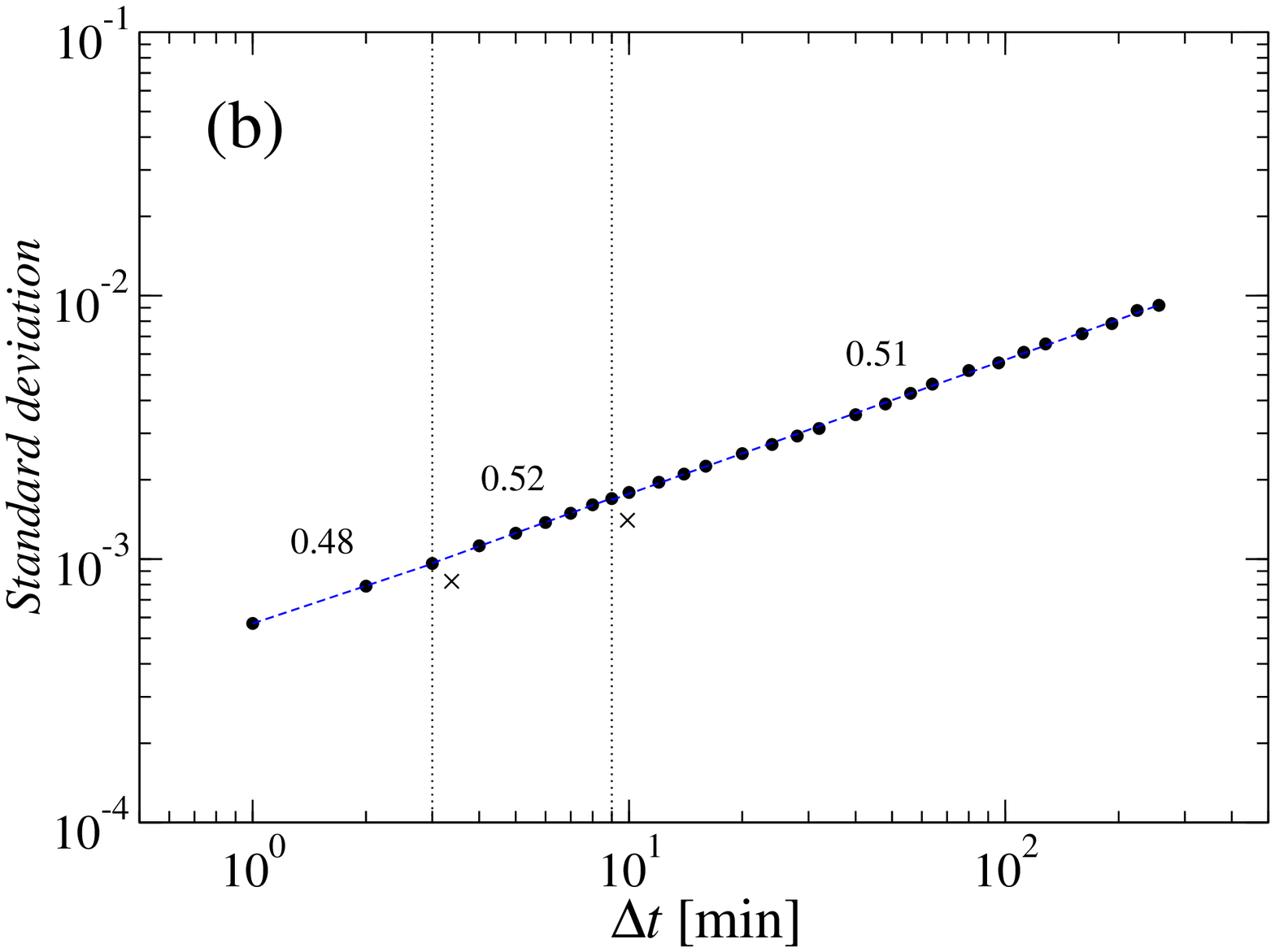}
\caption{Standard deviation of WIG20 fluctuations for different time scales ranging from 
$\Delta t=1$ to $\Delta t=256$ minutes. Time series of returns were divided into two parts of 
approximately equal length covering older (a) and newer (b) time interval. Dashed lines 
represent fits in terms of $\sigma(\Delta t) \sim \Delta t^{\delta}$. Vertical 
lines indicate crossover points ($\times$) between different diffusion regimes.} 
\end{figure}

\section{Summary}

The main conclusion to be drawn from the analysis presented above is that the formalism of 
nonextensive statistical mechanics based on the generalised Tsallis entropy seems to offer 
the best theoretical framework to compactly quantify the probability distributions of the 
stock market fluctuations on various time scales. This observation may appear helpful in 
formulating a generalised CLT when the summed stochastic variables - like the financial once 
- are not completely uncorrelated. It also gives further support to the option 
pricing~\cite{Black,Merton} models based on the $q$-Gaussians~\cite{Borland}. This analysis 
has been performed for the Warsaw Stock Market WIG20 high frequency recordings. Even though 
commonly considered as emerging this market displays fluctuation and correlation 
characteristics typical for the mature markets. The returns distribution for the small time 
lags conforms well to the inverse cubic power law. However, the departure from this law when 
increasing the time lags is seen to be faster than for any market presented so far in the 
literature. None of those previous studies for other markets refers however to such a recent 
time period and it is likely that going from past to present effectively contracts the time 
scales characteristic to the internal market dynamics and a similar effect may be seen for 
the other markets as well. In fact a trace of such an effect has already been noticed in 
ref.~\cite{Drozdz1}. This is an interesting issue for further study. We show that data from 
the Polish stock market exhibits long-range nonlinear correlations which can be seen in terms 
of the multifractality and the higher-order correlations in returns. Existence of such 
effects provides further arguments for considering the nonextensive statistical mechanics an 
inspiring framework for quantifying the stock market dynamics.

\end{document}